\documentclass[11pt,amsmath]{article}
\usepackage{graphicx}
\usepackage{amsmath}
\usepackage{amssymb}
\usepackage{color}
\usepackage{ulem}
\usepackage{url}
\usepackage{fancyvrb}
\usepackage{cleveref}
\usepackage{fancyhdr}
\usepackage[margin=1in]{geometry}
\usepackage{balance}
\usepackage{float}
\usepackage[english]{babel}
\usepackage{array}
\usepackage{dblfloatfix}
\usepackage{fixltx2e}
\usepackage[section]{placeins}
\usepackage{authblk}
\usepackage[font=footnotesize,labelfont=bf]{caption}
\usepackage[numbers,sort&compress,comma]{natbib}

\usepackage{url}
\usepackage{cleveref}

\begin{document}

% \begin{centering}
% \noindent\Huge{\textbf{Shape Allophiles Improve Entropic Assembly}}
\title{\textbf{Shape Allophiles Improve Entropic Assembly}}

% \noindent\Large{\textbf{Eric S. Harper,\textit{$^{a}$} Ryan L. Marson,\textit{$^{a}$} Joshua A. Anderson,\textit{$^{b}$} Greg van Anders,\textit{$^{b}$} and
% Sharon C. Glotzer\textit{$^{a,b,c}$}}} \\
% \author{\textbf{Eric S. Harper,\textit{$^{a}$} Ryan L. Marson,\textit{$^{a}$} Joshua A. Anderson,\textit{$^{b}$} Greg van Anders,\textit{$^{b}$} and Sharon C. Glotzer\textit{$^{a,b,c}$}}}
\author[1]{Eric S. Harper}
\author[1]{Ryan L. Marson}
\author[2]{Joshua A. Anderson}
\author[2]{Greg van Anders}
\author[1,2,3]{Sharon C. Glotzer}
\affil[1]{Department of Materials Science and Engineering, University of Michigan, Ann Arbor, Michigan 48109, USA.}
\affil[2]{Department of Chemical Engineering, University of Michigan, Ann Arbor, Michigan 48109, USA.}
\affil[3]{Biointerfaces Institute, University of Michigan, Ann Arbor, Michigan 48109, USA. E-mail: sglotzer@umich.edu}
\maketitle
% \noindent\large{\textit{$^{a}$~Department of Materials Science and Engineering, University of Michigan, Ann Arbor, Michigan 48109, USA.}}\\
% \noindent\large{\textit{$^{b}$~Department of Chemical Engineering, University of Michigan, Ann Arbor, Michigan 48109, USA.}}\\
% \noindent\large{\textit{$^{c}$~Biointerfaces Institute, University of Michigan, Ann Arbor, Michigan 48109, USA. E-mail: sglotzer@umich.edu}}\\
% \end{centering}

\begin{abstract}
\noindent
We investigate a class of ``shape allophiles'' that fit together like puzzle pieces as a method to access and stabilize desired structures by controlling directional entropic forces. Squares are cut into rectangular halves, which are shaped in an allophilic manner with the goal of re-assembling the squares while self-assembling the square lattice. We examine the assembly characteristics of this system via the potential of mean force and torque, and the fraction of particles that entropically bind. We generalize our findings and apply them to self-assemble triangles into a square lattice via allophilic shaping. Through these studies we show how shape allophiles can be useful in assembling and stabilizing desired phases with appropriate allophilic design.
\end{abstract}

\section*{Introduction}

Self-assembly and allophily both play crucial roles in biological systems\cite{King2012,Lai2012,Lanci2012,Bortolini2014}. The way in which proteins, enzymes, and DNA fit together have inspired and guided researchers to consider ``allophilic'' geometries\cite{Docking2003,Piazza2004,King2012,Murray2013,Ulijn2015}, so-called because these geometries are specifically designed to ``like'' each other. These structures fit together like puzzle pieces with the ability to create hierarchical structures. Lock-and-key colloids\cite{Konig2008,Sacanna2010,Odriozola2013,Wang2014,Laura2015,Ahmed2015}, designed and synthesized to exploit shape and entropic depletion forces for assembly, are an example of shape allophiles. Of course, biological systems such as enzymes and proteins rely not only on geometry and entropy, but also on intra- and inter-molecular forces to guide and hold structures in place. It remains to untangle this relationship and understand the contribution of the entropic interaction, answering the question: is shape alone enough to assemble these structures?

Many ordered phases can be obtained via shape alone \cite{Padilla2001,Damasceno2012}, arising from entropic patches\cite{vanAnders2014,Anders2014} and their associated directional entropic forces (DEFs) \cite{vanAnders2014,Millan2014,Damasceno}. DEFs are emergent, resulting from the collective behavior of the system. Much like enthalpic patches\cite{Zhang2004,Glotzer2007,Kraft2011,Kraft2012,Doppelbauer2012}, these entropic patches allow for particles to form entropic bonds with neighboring particles\cite{vanAnders2014,Anders2014}.

Here we show that shape allophilic patterning is a quantifiable and designable approach that can be used to increase the strength of the entropic bond formed between neighboring pairs of particles. This results in a larger number of bonded particles and a higher degree of order of the self-assembled phase. We demonstrate these findings on a system of rectangular colloids, and explain our findings by calculating and comparing the potential of mean force and torque (PMFT)\cite{vanAnders2014,Anders2014} as a measure of the strength and directionality of the DEFs between particles, as well as a bonding metric to quantify the assembly propensity and success of specific shape-allophilic patterns. Finally we apply this knowledge to allophilically pattern and induce an ordered self-assembled phase of triangles that is otherwise inaccessible.

\section*{Methods}

\FloatBarrier
\begin{figure}[htbp!]
\begin{center}
\includegraphics{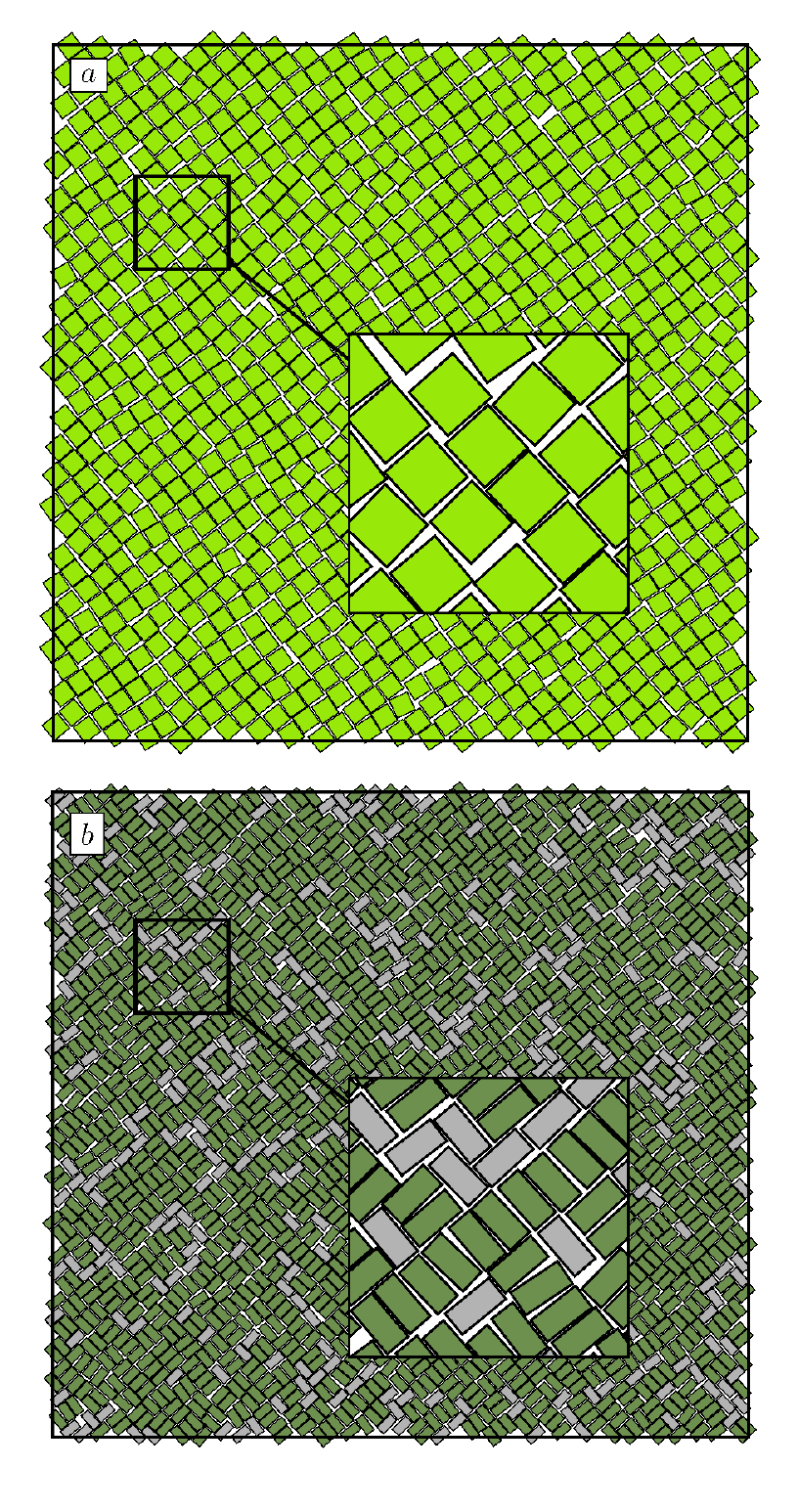}
\caption{Self-assembly of \textit{(a)} $N=1024$ squares and \textit{(b)} $N=2048$ $2:1$ aspect ratio rectangles. The rectangles do not successfully assemble the square lattice, but rather the similar yet distinct random domino (parquet) tiling due to the mixing entropy\cite{Donev2006} available to the rectangles.}
\label{fig:SquareParquet}
\end{center}
\end{figure}

\FloatBarrier
\begin{figure}[htbp!]
\begin{center}
\includegraphics{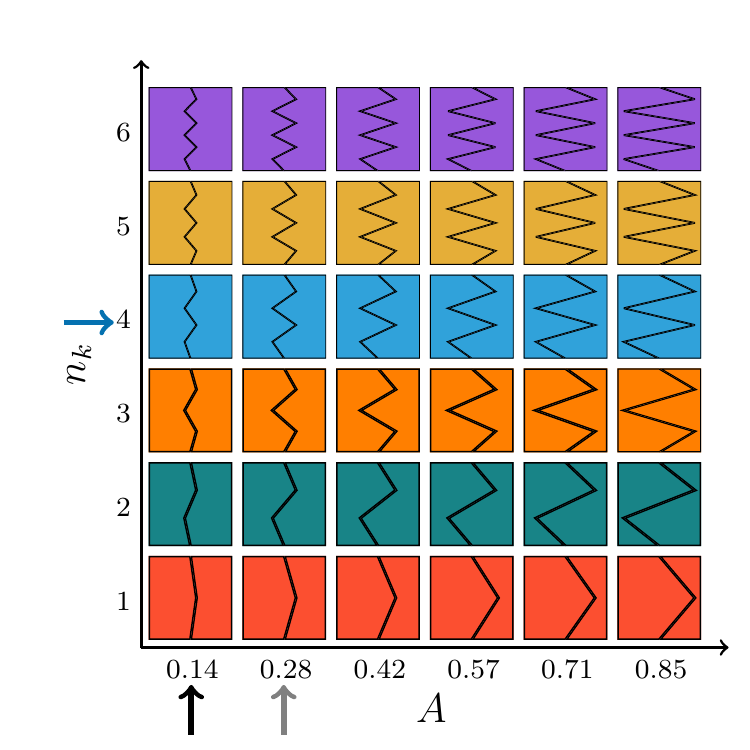}
\caption{Phase space schematic of ``parent'' squares cut into rectangles with triangle wave allophilic shaping. Amplitude, $A$, and wavenumber, $n_k$, are used to manipulate the patterning. Colored arrows mark the cuts through phase space shown in Figure~\ref{fig:4x4Combo}.}
\label{fig:ShapeVars}
\end{center}
\end{figure}

We investigate a simple two-dimensional system to determine the effect that allophilic shaping has on assembly propensity. We choose squares as a generic class of ``parent'' shape to determine if their ``child'' halves would reassemble the parent square and self-assemble the square lattice, as shown in Figure~\ref{fig:SquareParquet}. Squares and rounded squares have been previously investigated, providing a solid foundation for further study\cite{Wojciechowski2004,Smallenburg2012,Avendano2012,Zhao2011}. The choice of child shape for this study is $2:1$ aspect ratio rectangles, selected due to their tendency to self-assemble the random domino (parquet) tiling\cite{Donev2006,John2005,Triplett2008} instead of the square lattice because of the higher entropy of the former, shown in Figure~\ref{fig:SquareParquet}\textit{(b)}. We perform a thorough analysis of this system, and provide design rules to predict the assembly behavior of another polygonal system.

We employ triangle waves to form the allophilic shaping with amplitude, $A$, and wavenumber, $n_k$ (Figure~\ref{fig:ShapeVars}) applied to rectangles of long-edge length $L$. Possible amplitudes are $A \in \left[0,0.5L\right]$, which are reported as a fraction of $0.5L$: $A \in [0,1]$. We choose positive integer values for $n_k$ such that $n_k=1$ corresponds to one half-wavelength. A rectangle corresponds to $A=0, \; n_k=0$.

We use HPMC\cite{Anderson2013,Anderson2015}, an in-house HOOMD-Blue\cite{hoomd,Anderson2008} plugin for hard particle Monte Carlo (MC), to perform simulations of $N=10082$ particles ($5041$ pairs) on $32$ CPUs on XSEDE\cite{XSEDE} Stampede\cite{stampede}. HPMC utilizes MPI domain decomposition, based on the implementation\cite{Glaser2015} in HOOMD-Blue. Simulations are performed in the $NPT$ thermodynamic ensemble, allowing the simulation box to shear. A step size of $\Delta \left( \beta L^2 P \right) = 0.2, \; L=\sqrt{2}$ is used to scan through the system, identifying the highest density fluid and lowest density solid via the rate of decay in the orientation correlation function\cite{Dillmann2012,Bernard2011,Deutschlander2013}, as detailed in the SI$^\dag$. The system starts with an artificially constructed low-density crystal $\left( \phi=0.2 \right)$ for convenience, and is thermalized to a random configuration before compressing to a higher density fluid. The system is then compressed to the target pressure, allowing the system to run $4 \times 10^6$ sweeps at each incremental increase in pressure. Once at the target pressure, the system runs to equilibration in the target crystal phase, for at least $4 \times 10^7$ Monte Carlo sweeps.

We hypothesize that allophilic shaping through the introduction of allophilic faceting will form entropic patches, which will increase the DEFs, as the DEFs are correlated with facet size\cite{vanAnders2014}. The PMFT allows for the quantitative measurement and comparison of DEFs \textit{via} the calculation of the free energy associated with the pair correlation function. PMFTs are measured using the highest-density fluids found in the phase space sweep. Allophilic shaping must be able to overcome the random domino phase, entropically favored by unpatterned rectangles, by encouraging the formation of the parent squares and discouraging the formation of local motifs contributing to the random domino tiling.

The assembly propensity resulting from this shaping is measured by calculating the fraction of bonded particles, $f_b$, in the lowest density solid found in the phase space sweep. Two particles are considered bonded when they are within a given distance of each other, $d$, and the difference in angle between the inter-particle vectors and the perfectly bonded inter-particle vectors is within a tolerance $\theta \approx 0.1 \pi$. This allows for the measurement and evaluation of the assembly propensity of the allophilic shaping to form the desired parent structure. The fraction of bonded particles is $f_b=\frac{N_b}{N}$, where $N_b$ is the number of particles that are bonded. For rectangles without allophilic shaping, $f_b$ is halved to account for the two possible binding surfaces. The amplitude, $A$, is held constant while wavenumber, $n_k$, is varied to determine its effect on entropic bond strength and assembly propensity; the reverse is done to determine the effect of $n_k$.

\section*{Results and Discussion}

\FloatBarrier
\begin{figure}[htbp!]
\begin{center}
\includegraphics{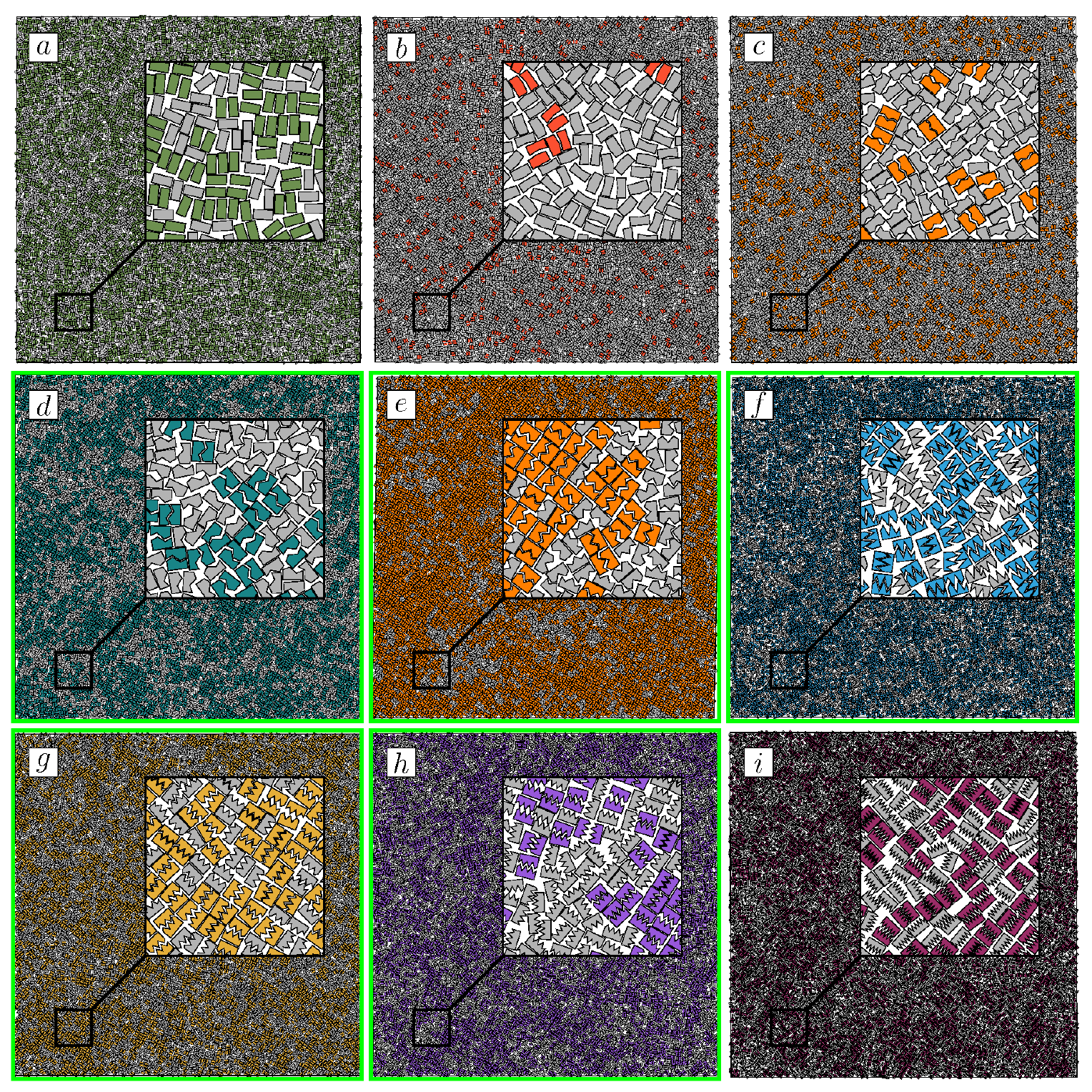}
\caption{Snapshots of equilibrium simulation frames of the lowest density solid found for \textit{(a)} rectangles $\left(\beta L^2 P=9.0\right)$ and allophilic rectangles with \textit{(b)} $n_k=1, \; A=0.14, \; \beta L^2 P=10.0$, \textit{(c)} $n_k=3, \; A=0.14, \; \beta L^2 P=11.8$, \textit{(d)} $n_k=2, \; A=0.28, \; \beta L^2 P=12.6$, \textit{(e)} $n_k=3, \; A=0.28, \; \beta L^2 P=13.2$, \textit{(f)} $n_k=4, \; A=0.57, \; \beta L^2 P=8.0$, \textit{(g)} $n_k=5, \; A=0.28, \; \beta L^2 P=11.2$, \textit{(h)} $n_k=6, \; A=0.28, \; \beta L^2 P=9.8$, \textit{(i)} $n_k=10, \; A=0.28, \; \beta L^2 P=9.2$. Wavenumbers increase left to right, while amplitude increases top to bottom. Bonded shapes are colored as in Figure~\ref{fig:ShapeVars}; otherwise, they are colored grey. Shapes that improve assembly of the square lattice relative to rectangles are indicated by a green surrounding box. Values for $f_b$ can be found in Figure~\ref{fig:4x4Combo}\textit{(d-i)}, while values for the pressure of the lowest density solids, $\beta L^2 P^*$, can also be found in Figure~\ref{fig:4x4Combo}\textit{(j-l)}.}
\label{fig:AssemblyArray}
\end{center}
\end{figure}

\FloatBarrier
\begin{figure}[htbp!]
\begin{center}
\includegraphics{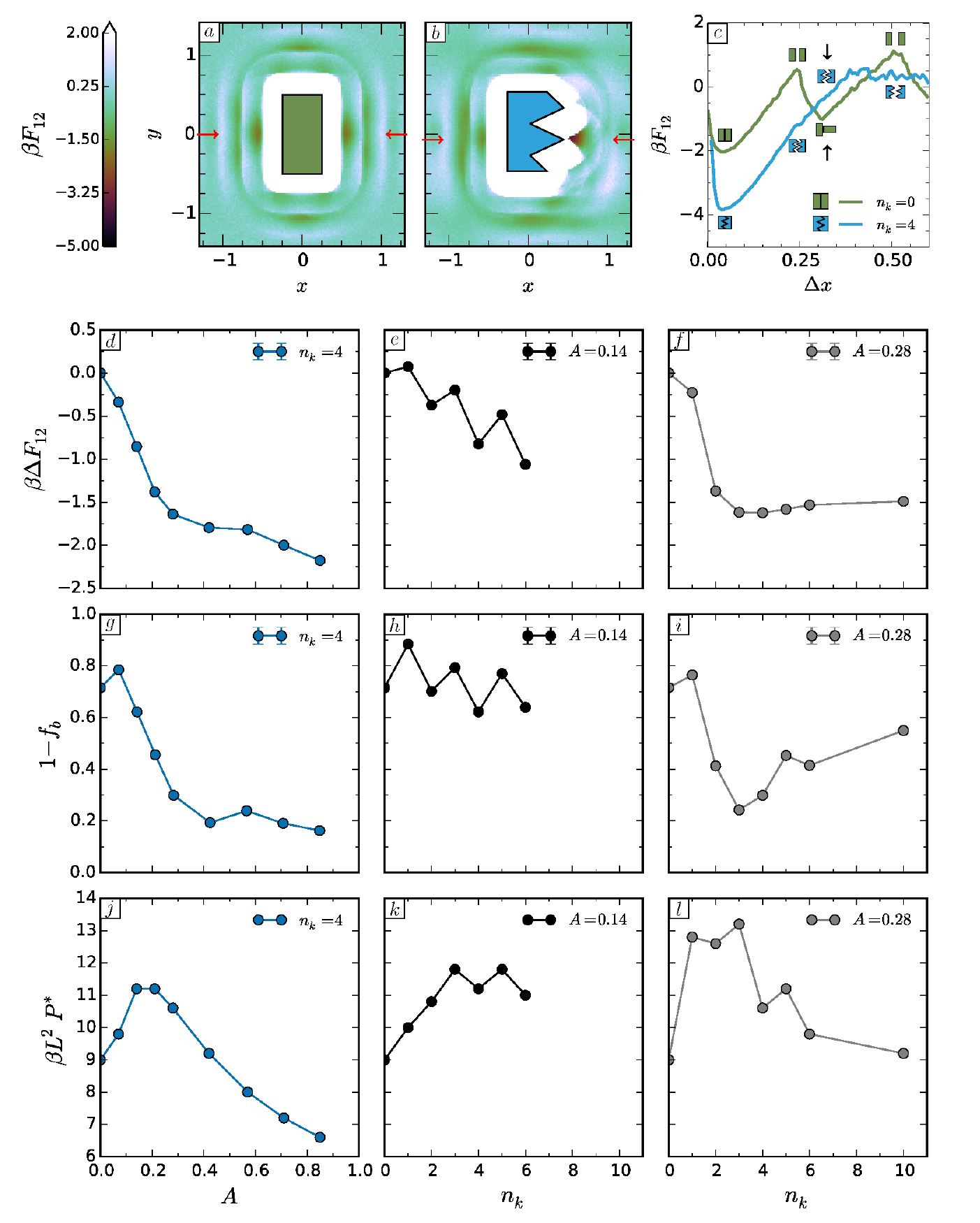}
\caption{PMFTs for \textit{(a)} rectangles and \textit{(b)} allophilic rectangles of $n_k=4$, $A=0.42$. \textit{(c)} Cross-section through binding well of PMFTs shown in \textit{(a, b)} (indicated with red arrows) only including the first neighbor shell. For ease of comparison, $\Delta x$ is the distance from the fully bonded position. Schematics of the local configuration are inset to aid in the interpretation of the PMFT features. Note that the allophilic shaping is effective at removing the minima associated with the random domino tiling indicated with black arrows at $\left( \Delta x \approx 0.3 \right)$. Difference in the bonding well energies $\beta \Delta F_{12}$ (upper bound) for constant \textit{(d)} $n_k=4$ with varying $A$, and for constant \textit{(e)} $A=0.14$, \textit{(f)} $A=0.28$ for varying $n_k$. Average fraction of unbonded particles $1-f_b$ (lower bound) for constant \textit{(g)} $n_k=4$ with varying $A$, and for constant \textit{(h)} $A=0.14$,  \textit{(i)} $A=0.28$ with varying $n_k$. Pressures $\beta L^2 P^*$ at which lowest density solid is found (upper bound) for constant \textit{(j)} $n_k=4$ with varying $A$, and for constant \textit{(k)} $A=0.14$, \textit{(l)} $A=0.28$ for varying $n_k$. \textit{(d,g)} show a direct correlation between the increase in DEFs and fraction of particles that bind $\left( f_b \right)$ due to increase in $A$, while \textit{(e, f, h, i)} show the threshold-like nature of $A$ for increasing the DEFs, and that there exists an optimal value of $n_k$. Error bars are reported for \textit{(d - i)} as the standard deviation of the average of $4$ replicate simulations at the same state point. Error for \textit{(d, e, f)} is $O \left( 0.005 \right)$, while error for \textit{(g, h, i)} is $O \left( 0.0005 \right)$, which is smaller than the markers used. No error bars reported for \textit{(j, k, l)} as pressure is an independent variable in the simulation.}
\label{fig:4x4Combo}
\end{center}
\end{figure}

Figure~\ref{fig:AssemblyArray} shows that in the lowest density solid, suitable choices of the amplitude ($A$) and wavenumber ($n_k$) lead to a substantial increase in the local ordering of the system into the desired local square motif. In SI$\dag$ Figure $5$, the images in Figure~\ref{fig:AssemblyArray} are replaced by the best achieved thermodynamic assembly. To isolate the effect allophilic shaping has on the strength of the entropic bond, as measured by the depth of the primary bonding well, cross-sections of the PMFTs are computed through this well, shown in Figure~\ref{fig:4x4Combo}. The difference in the depth of this well and the binding well of a rectangle, $\beta \Delta F_{12} = \beta F_{12} - \beta F_{12,\text{rectangle}}$, is calculated. We observe that amplitude has the strongest effect on the depth of the well, producing a difference in free energy of $\beta \Delta F_{12} > 2$ at the highest amplitudes investigated (Figure~\ref{fig:4x4Combo}\textit{(d)}). This indicates that the strength of the entropic bond increases with increasing amplitude. This increased entropic bond strength favors the desired square phase and disfavors the similar yet distinct random domino tiling, as seen by the merging of the first two minima with increasing amplitude, shown in Figure~\ref{fig:4x4Combo} and the SI$\dag$.

The wavenumber affects the depth of the binding well to a lesser extent than does the amplitude. At lower amplitudes, $A=0.14$ (Figure~\ref{fig:4x4Combo}\textit{(e)}), the difference in free energy of $\beta \Delta F_{12} < 1$ is not as significant as the gain in mixing entropy: even-valued $n_k$ have $N$ particles with which to bind, while odd-valued $n_k$ have only $\frac{N}{2}$ particles with which to bind. As even and odd wavenumbers affect the depth of the bonding well differently, the entropy associated with having a homogenous \textit{versus} a heterogeneous system dominates the DEFs created from allophilic shaping\cite{Schultz2015}. At slightly higher amplitudes, $A=0.28$, (Figure~\ref{fig:4x4Combo}\textit{(f)}), the effect is significant enough to overcome this mixing entropy because even and odd wavenumbers no longer have different effects on the depth of the well. At higher amplitudes, both even and odd wavenumbers are effective in deepening the primary well, except for $n_k=1$. At fixed $A=0.28$, increasing the wavenumber above $n_k=4$ leads to a slight \textit{decrease} in the depth of the primary well, suggesting that the optimal wavenumber is $n_k = 3\text{ or }4$ (Figure~\ref{fig:4x4Combo}\textit{(f)}).

\FloatBarrier
\begin{figure}[htbp]
\begin{center}
\includegraphics{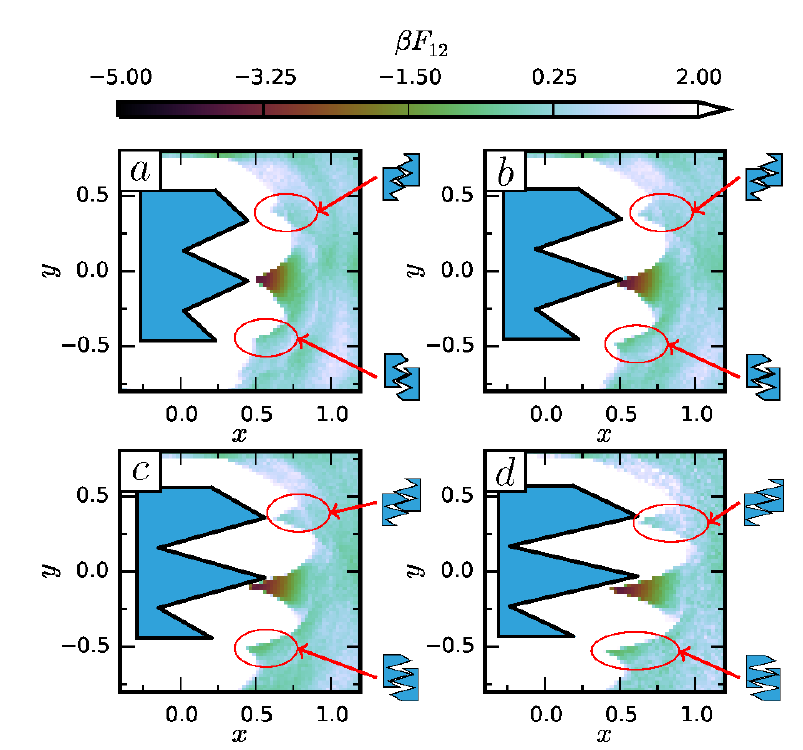}
\caption{PMFTs for \textit{(a)} $A=0.42$, \textit{(b)} $A=0.57$, \textit{(c)} $A=0.71$, and \textit{(d)} $A=0.85$ at constant $n_k=4$. As $A$ increases, secondary binding wells begin to emerge, circled in red (especially evident in \textit{(c, d)}), indicating that the alternate binding configurations are probable enough to prevent a further increase in $f_b$ that would be expected given the increase in the depth of the bonding well (Figure~\ref{fig:4x4Combo}\textit{(d, g)})}
\label{fig:pmftnk4}
\end{center}
\end{figure}

\FloatBarrier
\begin{figure}[htbp]
\begin{center}
\includegraphics{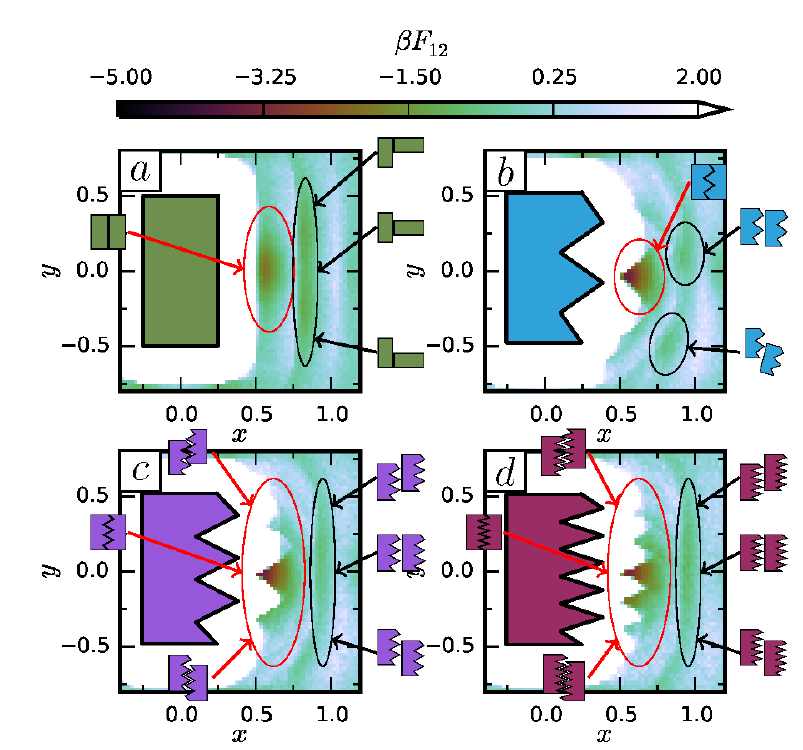}
\caption{PMFTs for \textit{(a)} rectangles and allophilic rectangles with \textit{(b)} $n_k=4$, \textit{(c)} $n_k=6$, and \textit{(d)} $n_k=10$ at constant $A=0.28$. The primary (and secondary in \textit{(c, d)}) wells are circled in red, while the wells corresponding to the random domino tiling are circled in black. The emergence of secondary binding wells in \textit{(c, d)} compared to the singular well in \textit{(b)} indicate that the allophilic shaping is effective at binding the shapes together, but not in selecting the desired configuration. The re-merging of the wells associated with unbonded particles in \textit{(c, d)} indicate that the increase in $n_k$ results in the allophilic patch appearing flatter to the unpatterned side of the shape, reducing its effectiveness in selecting for the entropic bond}
\label{fig:pmftnk10}
\end{center}
\end{figure}

The fraction of bonded particles, $f_b$, in the lowest density solid found serves as a measure of the assembly propensity of the target crystal. Rectangles self-assemble the random domino tiling, while allophilic rectangles are better able to self-assemble the square lattice as seen in Figure~\ref{fig:AssemblyArray,fig:4x4Combo}. This effect is more pronounced at higher values of $A$ and $n_k>1$. For both $A$ and $n_k$, we observed an initial decrease in pairing, which suggests that small amounts of shaping do not increase the DEFs enough to overcome the entropic repulsion caused by vertices\cite{vanAnders2014}. The fraction $f_b$ plateaus (Figure~\ref{fig:4x4Combo}\textit{(g)}) at $A \approx 0.4$, while for $n_k$ at constant $A=0.28$ there is a maximum in $f_b$ (shown as a minimum in $1-f_b$) at $n_k=3$ (Figure~\ref{fig:4x4Combo}\textit{(i)}).

As $A$ increases, the width of the bonding well increases (Figure~\ref{fig:4x4Combo}\textit{(f)} and SI$\dag$), eventually eliminating the random domino bonding well, and disallowing this competing motif. However, the effect that the increased $A$ has on pairing eventually plateaus (Figure~\ref{fig:4x4Combo}\textit{(g)}). As seen in Figure~\ref{fig:pmftnk4}, the PMFT landscape shows the emergence of misbonded particles, introducing competing motifs that are favorable enough to inhibit further improvement of the square lattice.

Interestingly, even and odd $n_k$ do not seem to impact the particle pairing once the threshold value for $A$ is met. As discussed previously, odd $n_k$ have only $\frac{N}{2}$ particles with which to bind; thus, we expected that $f_b$ for odd values of $n_k$ would be less than that for even numbers, for which any particle is a correct match, as seen for $A=0.14$ (Figure~\ref{fig:4x4Combo}\textit{(e)})\cite{Schultz2015}. Instead, we see that once the DEFs are strong enough, the difference in fraction of available particles with which to bind no longer matters; in fact at $A=0.28$ the odd $n_k=3$ formed the most pairs (Figure~\ref{fig:4x4Combo}\textit{(i)}). This unexpectedly high assembly propensity demonstrates how successful allotropic shaping is at forming the desired bond using a relatively small amount of shaping.

As $n_k$ increases past $n_k=4$, the fraction of bonded particles decreases (Figure~\ref{fig:4x4Combo}\textit{(i)}). As seen in Figure~\ref{fig:pmftnk10}, the additional teeth allow for undesired bonds to form between the shapes. Additionally, the re-appearance of the well corresponding to unbonded particles as seen in Figure~\ref{fig:pmftnk10} leads to the conclusion that the teeth are close enough to sterically mimic a flat face, thus allowing for a parquet-like phase to still be entropically competitive. This study also shows that allophilic shaping successfully discourages the formation of local motifs that lead to the random domino tiling. Allophilic particles simply cannot be in a ``T'' configuration without wasting too much available inter-tooth space.

Allophilic shaping is a useful shaping technique to avoid alternate motifs that prevent the formation of the parent shape. Collectively, our results in Figure~\ref{fig:AssemblyArray}\textit{(g,i)} suggest a simple design rule for allophilic patterning that promotes the formation of local motifs consistent with the desired global motif: the particle features that result from allophilic shape patterning should be on the order of $A \approx 0.3$ which corresponds to $15\%$ of the particle size and $3-4$ in number. In terms of amplitude, smaller features do not lead to sufficiently strong entropic binding to overcome mixing entropy. In terms of wavenumber, fewer features do not provide sufficient selectivity for the entropic bonds while more features begin to introduce competing motifs.

\FloatBarrier
\begin{figure}[htbp!]
\begin{center}
\includegraphics{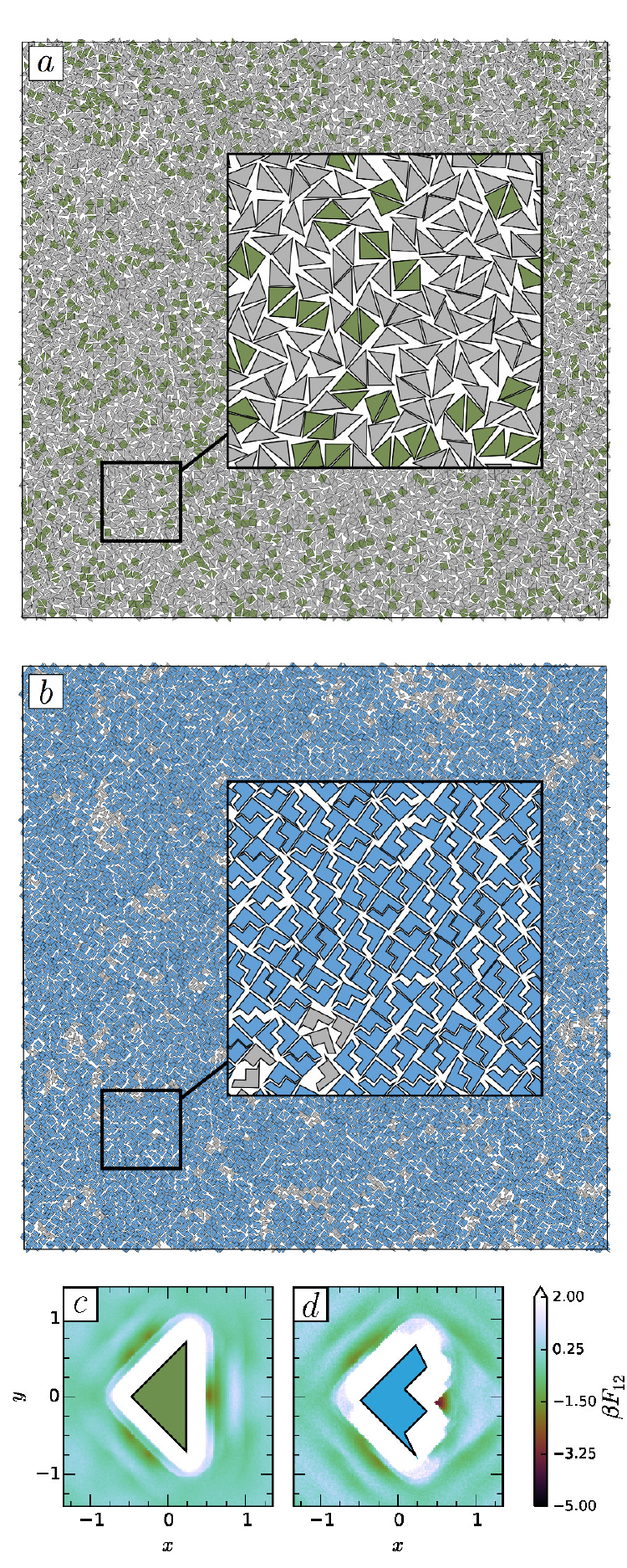}
\caption{Self assembly of \textit{(a)} right-isosceles triangles and \textit{(b)} allophilic triangles with $n_k=4, \; A=0.28$ showing that allophilic interactions from entropic patterning leads to increased bonding. Bonded shapes are colored \textit{(a)} green and \textit{(b)} blue; otherwise, they are colored grey. PMFTs of \textit{(c)} right-isosceles triangles and \textit{(d)} allophilic triangles with $n_k=4, \; A=0.28$}
\label{fig:TriAssembly}
\end{center}
\end{figure}

\balance

We show the efficacy of these design rules using a system of right-isosceles triangles. Right-isosceles triangles do not self-assemble the square phase or any ordered phase, as seen in Figure~\ref{fig:TriAssembly}\textit{(a)}, in contrast to the rectangular system previously studied. Triangle-wave shaping is applied in the same fashion as for rectangles to right-isosceles triangles. Intermediate values of $n_k=4, \; A=0.28$ are utilized to demonstrate that an experimentally feasible level of shaping is capable of assembling the desired phase. We find these values of $n_k=4, \; A=0.28$ are successful at inducing ordering in the system, as measured by the pair fraction of $f > 0.95$.

As seen in Figure~\ref{fig:TriAssembly}\textit{(c, d)}, the right isosceles triangle system has potential wells analogous to those in the random domino tiling in rectangles, which now corresponds to a configuration where the base and hypotenuse align. This arrangement is not compatible with the desired square lattice. And unlike rectangles, there is no alternate tiling available, causing a disordered solid phase to form rather than the desired crystal. By applying the allophilic shaping, the hypotenuse-hypotenuse bond is strengthened while eliminating the base-hypotenuse bond, allowing for the self-assembly of the desired phase, more effectively than allophilic rectangles ($f_{b,\text{triangle}}\approx 0.95$, $f_{b,\text{rect},\text{max}}\approx 0.85$).

\section*{Conclusion}

We have shown how allophilic shaping improves the self-assembly of desired phases without the use of DNA functionalization, external fields, or other types of intrinsic inter-particle interactions. These entropic interactions both stabilize desired phases and form phases that are otherwise unobtainable. PMFT analysis shows that allophilic shaping is able to increase the DEFs between particles, and that the strength of this force is tunable by varying the shaping. The entropic patch\cite{vanAnders2014} created via allophilic shaping adds another method to the growing number of self-assembly tools available to researchers. In the self-assembly of square-derived polygons, the order of the assembly was increased by allophilic interaction; in fact, an ordered phase was achieved with allophilic right-isosceles triangles where unpatterned triangles failed. We conjecture that in nature, where crowded, non-convex geometries are common (\textit{e.g.} proteins), shape allophilic, entropic  interactions contribute significantly to their lock-and-key binding. Our results from a simple, model two-dimensional system suggest that features need only be $\approx 15\%$ of particle size to get the desired binding without enthalpic interactions. It would be of interest to test this in a more complex three-dimensional system. Here we studied a monovalent shape allophile; future work should also investigate different types of allophilic shaping, including multivalent shape allophiles.

Finally, we note that prior work for convex hard particles has shown that vertices are effectively entropically repulsive \cite{Anders2014,vanAnders2014,Gantapara2015}, but vertex proximity  has been shown to be important for the plasmonic response of anisotropic particles \cite{Gao2013,Rosen2014}. Our allophilic particle yield motifs with relatively high vertex proximity, which are potentially useful for improving plasmonic response in systems of anisotropic particles.

\section*{Acknowledgements}

We thank Richmond Newman and Sari van Anders for many useful discussions. This research used the Extreme Science and Engineering Discovery Environment (XSEDE), which is supported by National Science Foundation grant number ACI-1053575 (XSEDE award DMR 140129), and was also supported through computational resources and services provided by Advanced Research Computing at the University of Michigan, Ann Arbor. This material is based upon work supported in part by the U.S. Army Research Office under Grant Award No. W911NF-10-1-0518 and also by the DOD/ASD(R\&E) under Award No. N00244-09-1-0062. Any opinions, findings, and conclusions or recommendations expressed in this publication are those of the author(s) and do not necessarily reflect the views of the DOD/ASD(R\&E). E.H. also supported by the National Science Foundation DGE 0903629 Open Data IGERT. J.A. also supported by the National Science Foundation, Division of Materials Research Award \# DMR 1409620. R.M. acknowledges support from the University of Michigan Rackham Merit Fellowship program.

%%%REFERENCES%%%
\bibliography{arXiv.bbl}

\begin{thebibliography}{46}%
\makeatletter
\providecommand \@ifxundefined [1]{%
 \@ifx{#1\undefined}
}%
\providecommand \@ifnum [1]{%
 \ifnum #1\expandafter \@firstoftwo
 \else \expandafter \@secondoftwo
 \fi
}%
\providecommand \@ifx [1]{%
 \ifx #1\expandafter \@firstoftwo
 \else \expandafter \@secondoftwo
 \fi
}%
\providecommand \natexlab [1]{#1}%
\providecommand \enquote  [1]{``#1''}%
\providecommand \bibnamefont  [1]{#1}%
\providecommand \bibfnamefont [1]{#1}%
\providecommand \citenamefont [1]{#1}%
\providecommand \href@noop [0]{\@secondoftwo}%
\providecommand \href [0]{\begingroup \@sanitize@url \@href}%
\providecommand \@href[1]{\@@startlink{#1}\@@href}%
\providecommand \@@href[1]{\endgroup#1\@@endlink}%
\providecommand \@sanitize@url [0]{\catcode `\\12\catcode `\$12\catcode
  `\&12\catcode `\#12\catcode `\^12\catcode `\_12\catcode `\%12\relax}%
\providecommand \@@startlink[1]{}%
\providecommand \@@endlink[0]{}%
\providecommand \url  [0]{\begingroup\@sanitize@url \@url }%
\providecommand \@url [1]{\endgroup\@href {#1}{\urlprefix }}%
\providecommand \urlprefix  [0]{URL }%
\providecommand \Eprint [0]{\href }%
\providecommand \doibase [0]{http://dx.doi.org/}%
\providecommand \selectlanguage [0]{\@gobble}%
\providecommand \bibinfo  [0]{\@secondoftwo}%
\providecommand \bibfield  [0]{\@secondoftwo}%
\providecommand \translation [1]{[#1]}%
\providecommand \BibitemOpen [0]{}%
\providecommand \bibitemStop [0]{}%
\providecommand \bibitemNoStop [0]{.\EOS\space}%
\providecommand \EOS [0]{\spacefactor3000\relax}%
\providecommand \BibitemShut  [1]{\csname bibitem#1\endcsname}%
\let\auto@bib@innerbib\@empty
%</preamble>
\bibitem [{\citenamefont {Bortolini}\ \emph {et~al.}(2014)\citenamefont
  {Bortolini}, \citenamefont {Liu}, \citenamefont {Gronewold}, \citenamefont
  {Wang}, \citenamefont {Besenbacher},\ and\ \citenamefont
  {Dong}}]{Bortolini2014}%
  \BibitemOpen
  \bibfield  {author} {\bibinfo {author} {\bibfnamefont {C.}~\bibnamefont
  {Bortolini}}, \bibinfo {author} {\bibfnamefont {L.}~\bibnamefont {Liu}},
  \bibinfo {author} {\bibfnamefont {T.~M.~A.}\ \bibnamefont {Gronewold}},
  \bibinfo {author} {\bibfnamefont {C.}~\bibnamefont {Wang}}, \bibinfo {author}
  {\bibfnamefont {F.}~\bibnamefont {Besenbacher}}, \ and\ \bibinfo {author}
  {\bibfnamefont {M.}~\bibnamefont {Dong}},\ }\href {\doibase
  10.1039/C4SM01065E} {\bibfield  {journal} {\bibinfo  {journal} {Soft Matter}\
  }\textbf {\bibinfo {volume} {10}},\ \bibinfo {pages} {5656} (\bibinfo {year}
  {2014})}\BibitemShut {NoStop}%
\bibitem [{\citenamefont {King}\ \emph {et~al.}(2012)\citenamefont {King},
  \citenamefont {Sheffler}, \citenamefont {Sawaya}, \citenamefont {Vollmar},
  \citenamefont {Sumida}, \citenamefont {Andr\'{e}}, \citenamefont {Gonen},
  \citenamefont {Yeates},\ and\ \citenamefont {Baker}}]{King2012}%
  \BibitemOpen
  \bibfield  {author} {\bibinfo {author} {\bibfnamefont {N.~P.}\ \bibnamefont
  {King}}, \bibinfo {author} {\bibfnamefont {W.}~\bibnamefont {Sheffler}},
  \bibinfo {author} {\bibfnamefont {M.~R.}\ \bibnamefont {Sawaya}}, \bibinfo
  {author} {\bibfnamefont {B.~S.}\ \bibnamefont {Vollmar}}, \bibinfo {author}
  {\bibfnamefont {J.~P.}\ \bibnamefont {Sumida}}, \bibinfo {author}
  {\bibfnamefont {I.}~\bibnamefont {Andr\'{e}}}, \bibinfo {author}
  {\bibfnamefont {T.}~\bibnamefont {Gonen}}, \bibinfo {author} {\bibfnamefont
  {T.~O.}\ \bibnamefont {Yeates}}, \ and\ \bibinfo {author} {\bibfnamefont
  {D.}~\bibnamefont {Baker}},\ }\href {\doibase 10.1126/science.1219364}
  {\bibfield  {journal} {\bibinfo  {journal} {Science}\ }\textbf {\bibinfo
  {volume} {336}},\ \bibinfo {pages} {1171} (\bibinfo {year} {2012})},\ \Eprint
  {http://arxiv.org/abs/http://www.sciencemag.org/content/336/6085/1171.full.pdf}
  {http://www.sciencemag.org/content/336/6085/1171.full.pdf} \BibitemShut
  {NoStop}%
\bibitem [{\citenamefont {Lai}\ \emph {et~al.}(2015)\citenamefont {Lai},
  \citenamefont {King},\ and\ \citenamefont {Yeates}}]{Lai2012}%
  \BibitemOpen
  \bibfield  {author} {\bibinfo {author} {\bibfnamefont {Y.-T.}\ \bibnamefont
  {Lai}}, \bibinfo {author} {\bibfnamefont {N.~P.}\ \bibnamefont {King}}, \
  and\ \bibinfo {author} {\bibfnamefont {T.~O.}\ \bibnamefont {Yeates}},\
  }\bibfield  {booktitle} {\emph {\bibinfo {booktitle} {Trends in Cell
  Biology}},\ }\href {\doibase 10.1016/j.tcb.2012.08.004} {\bibfield  {journal}
  {\bibinfo  {journal} {Trends in Cell Biology}\ }\textbf {\bibinfo {volume}
  {22}},\ \bibinfo {pages} {653} (\bibinfo {year} {2015})}\BibitemShut
  {NoStop}%
\bibitem [{\citenamefont {Lanci}\ \emph {et~al.}()\citenamefont {Lanci},
  \citenamefont {MacDermaid}, \citenamefont {Kang}, \citenamefont {Acharya},
  \citenamefont {North}, \citenamefont {Yang}, \citenamefont {Qiu},
  \citenamefont {DeGrado},\ and\ \citenamefont {Saven}}]{Lanci2012}%
  \BibitemOpen
  \bibfield  {author} {\bibinfo {author} {\bibfnamefont {C.~J.}\ \bibnamefont
  {Lanci}}, \bibinfo {author} {\bibfnamefont {C.~M.}\ \bibnamefont
  {MacDermaid}}, \bibinfo {author} {\bibfnamefont {S.-g.}\ \bibnamefont
  {Kang}}, \bibinfo {author} {\bibfnamefont {R.}~\bibnamefont {Acharya}},
  \bibinfo {author} {\bibfnamefont {B.}~\bibnamefont {North}}, \bibinfo
  {author} {\bibfnamefont {X.}~\bibnamefont {Yang}}, \bibinfo {author}
  {\bibfnamefont {X.~J.}\ \bibnamefont {Qiu}}, \bibinfo {author} {\bibfnamefont
  {W.~F.}\ \bibnamefont {DeGrado}}, \ and\ \bibinfo {author} {\bibfnamefont
  {J.~G.}\ \bibnamefont {Saven}},\ }\href@noop {} {\ }\BibitemShut {NoStop}%
\bibitem [{\citenamefont {Chen}\ and\ \citenamefont
  {Weng}(2003)}]{Docking2003}%
  \BibitemOpen
  \bibfield  {author} {\bibinfo {author} {\bibfnamefont {R.}~\bibnamefont
  {Chen}}\ and\ \bibinfo {author} {\bibfnamefont {Z.}~\bibnamefont {Weng}},\
  }\href {\doibase 10.1002/prot.10334} {\bibfield  {journal} {\bibinfo
  {journal} {Proteins: Structure, Function, and Bioinformatics}\ }\textbf
  {\bibinfo {volume} {51}},\ \bibinfo {pages} {397} (\bibinfo {year}
  {2003})}\BibitemShut {NoStop}%
\bibitem [{\citenamefont {Paik}\ and\ \citenamefont
  {Murray}(2013)}]{Murray2013}%
  \BibitemOpen
  \bibfield  {author} {\bibinfo {author} {\bibfnamefont {T.}~\bibnamefont
  {Paik}}\ and\ \bibinfo {author} {\bibfnamefont {C.~B.}\ \bibnamefont
  {Murray}},\ }\href {\doibase 10.1021/nl401370n} {\bibfield  {journal}
  {\bibinfo  {journal} {Nano Letters}\ }\textbf {\bibinfo {volume} {13}},\
  \bibinfo {pages} {2952} (\bibinfo {year} {2013})},\ \bibinfo {note} {pMID:
  23668826},\ \Eprint
  {http://arxiv.org/abs/http://dx.doi.org/10.1021/nl401370n}
  {http://dx.doi.org/10.1021/nl401370n} \BibitemShut {NoStop}%
\bibitem [{\citenamefont {Piazza}(2004)}]{Piazza2004}%
  \BibitemOpen
  \bibfield  {author} {\bibinfo {author} {\bibfnamefont {R.}~\bibnamefont
  {Piazza}},\ }\href {\doibase 10.1016/j.cocis.2004.01.008} {\bibfield
  {journal} {\bibinfo  {journal} {Current Opinion in Colloid \& Interface
  Science}\ }\textbf {\bibinfo {volume} {8}},\ \bibinfo {pages} {515} (\bibinfo
  {year} {2004})}\BibitemShut {NoStop}%
\bibitem [{\citenamefont {Ulijn}(2015)}]{Ulijn2015}%
  \BibitemOpen
  \bibfield  {author} {\bibinfo {author} {\bibfnamefont {R.~V.}\ \bibnamefont
  {Ulijn}},\ }\href {http://dx.doi.org/10.1038/nnano.2015.59} {\bibfield
  {journal} {\bibinfo  {journal} {Nat Nano}\ }\textbf {\bibinfo {volume}
  {10}},\ \bibinfo {pages} {295} (\bibinfo {year} {2015})}\BibitemShut
  {NoStop}%
\bibitem [{\citenamefont {Ahmed}\ \emph {et~al.}(2015)\citenamefont {Ahmed},
  \citenamefont {van Anders}, \citenamefont {Chen},\ and\ \citenamefont
  {Glotzer}}]{Ahmed2015}%
  \BibitemOpen
  \bibfield  {author} {\bibinfo {author} {\bibfnamefont {N.~K.}\ \bibnamefont
  {Ahmed}}, \bibinfo {author} {\bibfnamefont {G.}~\bibnamefont {van Anders}},
  \bibinfo {author} {\bibfnamefont {E.~R.}\ \bibnamefont {Chen}}, \ and\
  \bibinfo {author} {\bibfnamefont {S.~C.}\ \bibnamefont {Glotzer}},\ }\href
  {http://arxiv.org/abs/1501.03130} {\bibfield  {journal} {\bibinfo  {journal}
  {arXiv}\ } (\bibinfo {year} {2015})}\BibitemShut {NoStop}%
\bibitem [{\citenamefont {Col\`{o}n-Mel\`{e}ndez}\ \emph
  {et~al.}(2015)\citenamefont {Col\`{o}n-Mel\`{e}ndez}, \citenamefont
  {Beltran-Villegas}, \citenamefont {van Anders}, \citenamefont {Liu},
  \citenamefont {Spellings}, \citenamefont {Sacanna}, \citenamefont {Pine},
  \citenamefont {Glotzer}, \citenamefont {Larson},\ and\ \citenamefont
  {Solomon}}]{Laura2015}%
  \BibitemOpen
  \bibfield  {author} {\bibinfo {author} {\bibfnamefont {L.}~\bibnamefont
  {Col\`{o}n-Mel\`{e}ndez}}, \bibinfo {author} {\bibfnamefont {D.~J.}\
  \bibnamefont {Beltran-Villegas}}, \bibinfo {author} {\bibfnamefont
  {G.}~\bibnamefont {van Anders}}, \bibinfo {author} {\bibfnamefont
  {J.}~\bibnamefont {Liu}}, \bibinfo {author} {\bibfnamefont {M.}~\bibnamefont
  {Spellings}}, \bibinfo {author} {\bibfnamefont {S.}~\bibnamefont {Sacanna}},
  \bibinfo {author} {\bibfnamefont {D.~J.}\ \bibnamefont {Pine}}, \bibinfo
  {author} {\bibfnamefont {S.~C.}\ \bibnamefont {Glotzer}}, \bibinfo {author}
  {\bibfnamefont {R.~G.}\ \bibnamefont {Larson}}, \ and\ \bibinfo {author}
  {\bibfnamefont {M.~J.}\ \bibnamefont {Solomon}},\ }\href {\doibase
  http://dx.doi.org/10.1063/1.4919299} {\bibfield  {journal} {\bibinfo
  {journal} {The Journal of Chemical Physics}\ }\textbf {\bibinfo {volume}
  {142}},\ \bibinfo {eid} {174909} (\bibinfo {year} {2015})}\BibitemShut
  {NoStop}%
\bibitem [{\citenamefont {K\"onig}\ \emph {et~al.}(2008)\citenamefont
  {K\"onig}, \citenamefont {Roth},\ and\ \citenamefont {Dietrich}}]{Konig2008}%
  \BibitemOpen
  \bibfield  {author} {\bibinfo {author} {\bibfnamefont {P.-M.}\ \bibnamefont
  {K\"onig}}, \bibinfo {author} {\bibfnamefont {R.}~\bibnamefont {Roth}}, \
  and\ \bibinfo {author} {\bibfnamefont {S.}~\bibnamefont {Dietrich}},\ }\href
  {http://stacks.iop.org/0295-5075/84/i=6/a=68006} {\bibfield  {journal}
  {\bibinfo  {journal} {Europhys. Lett.}\ }\textbf {\bibinfo {volume} {84}},\
  \bibinfo {pages} {68006} (\bibinfo {year} {2008})}\BibitemShut {NoStop}%
\bibitem [{\citenamefont {Odriozola}\ and\ \citenamefont
  {Lozada-Cassou}(2013)}]{Odriozola2013}%
  \BibitemOpen
  \bibfield  {author} {\bibinfo {author} {\bibfnamefont {G.}~\bibnamefont
  {Odriozola}}\ and\ \bibinfo {author} {\bibfnamefont {M.}~\bibnamefont
  {Lozada-Cassou}},\ }\href {\doibase 10.1103/PhysRevLett.110.105701}
  {\bibfield  {journal} {\bibinfo  {journal} {Phys. Rev. Lett.}\ }\textbf
  {\bibinfo {volume} {110}},\ \bibinfo {pages} {105701} (\bibinfo {year}
  {2013})}\BibitemShut {NoStop}%
\bibitem [{\citenamefont {Sacanna}\ \emph {et~al.}(2010)\citenamefont
  {Sacanna}, \citenamefont {Irvine}, \citenamefont {Chaikin},\ and\
  \citenamefont {Pine}}]{Sacanna2010}%
  \BibitemOpen
  \bibfield  {author} {\bibinfo {author} {\bibfnamefont {S.}~\bibnamefont
  {Sacanna}}, \bibinfo {author} {\bibfnamefont {W.~T.~M.}\ \bibnamefont
  {Irvine}}, \bibinfo {author} {\bibfnamefont {P.~M.}\ \bibnamefont {Chaikin}},
  \ and\ \bibinfo {author} {\bibfnamefont {D.}~\bibnamefont {Pine}},\ }\href
  {\doibase 10.1038/nature08906} {\bibfield  {journal} {\bibinfo  {journal}
  {Nature}\ }\textbf {\bibinfo {volume} {464}},\ \bibinfo {pages} {575}
  (\bibinfo {year} {2010})}\BibitemShut {NoStop}%
\bibitem [{\citenamefont {Wang}\ \emph {et~al.}(2014)\citenamefont {Wang},
  \citenamefont {Wang}, \citenamefont {Zheng}, \citenamefont {Yi},
  \citenamefont {Sacanna}, \citenamefont {Pine},\ and\ \citenamefont
  {Weck}}]{Wang2014}%
  \BibitemOpen
  \bibfield  {author} {\bibinfo {author} {\bibfnamefont {Y.}~\bibnamefont
  {Wang}}, \bibinfo {author} {\bibfnamefont {Y.}~\bibnamefont {Wang}}, \bibinfo
  {author} {\bibfnamefont {X.}~\bibnamefont {Zheng}}, \bibinfo {author}
  {\bibfnamefont {G.~R.}\ \bibnamefont {Yi}}, \bibinfo {author} {\bibfnamefont
  {S.}~\bibnamefont {Sacanna}}, \bibinfo {author} {\bibfnamefont {D.~J.}\
  \bibnamefont {Pine}}, \ and\ \bibinfo {author} {\bibfnamefont
  {M.}~\bibnamefont {Weck}},\ }\href {\doibase 10.1021/ja502699p} {\bibfield
  {journal} {\bibinfo  {journal} {Journal of the American Chemical Society}\
  }\textbf {\bibinfo {volume} {136}},\ \bibinfo {pages} {6866} (\bibinfo {year}
  {2014})}\BibitemShut {NoStop}%
\bibitem [{\citenamefont {Damasceno}\ \emph
  {et~al.}(2012{\natexlab{a}})\citenamefont {Damasceno}, \citenamefont
  {Engel},\ and\ \citenamefont {Glotzer}}]{Damasceno2012}%
  \BibitemOpen
  \bibfield  {author} {\bibinfo {author} {\bibfnamefont {P.~F.}\ \bibnamefont
  {Damasceno}}, \bibinfo {author} {\bibfnamefont {M.}~\bibnamefont {Engel}}, \
  and\ \bibinfo {author} {\bibfnamefont {S.~C.}\ \bibnamefont {Glotzer}},\
  }\href {\doibase 10.1126/science.1220869} {\bibfield  {journal} {\bibinfo
  {journal} {Science}\ }\textbf {\bibinfo {volume} {337}},\ \bibinfo {pages}
  {453} (\bibinfo {year} {2012}{\natexlab{a}})},\ \Eprint
  {http://arxiv.org/abs/1202.2177} {1202.2177 [cond-mat.soft]} \BibitemShut
  {NoStop}%
\bibitem [{\citenamefont {Padilla}\ \emph {et~al.}(2001)\citenamefont
  {Padilla}, \citenamefont {Colovos},\ and\ \citenamefont
  {Yeates}}]{Padilla2001}%
  \BibitemOpen
  \bibfield  {author} {\bibinfo {author} {\bibfnamefont {J.~E.}\ \bibnamefont
  {Padilla}}, \bibinfo {author} {\bibfnamefont {C.}~\bibnamefont {Colovos}}, \
  and\ \bibinfo {author} {\bibfnamefont {T.~O.}\ \bibnamefont {Yeates}},\
  }\href {\doibase 10.1073/pnas.041614998} {\bibfield  {journal} {\bibinfo
  {journal} {Proc. Natl. Acad. Sci. U.S.A.}\ }\textbf {\bibinfo {volume}
  {98}},\ \bibinfo {pages} {2217} (\bibinfo {year} {2001})},\ \Eprint
  {http://arxiv.org/abs/http://www.pnas.org/content/98/5/2217.full.pdf}
  {http://www.pnas.org/content/98/5/2217.full.pdf} \BibitemShut {NoStop}%
\bibitem [{\citenamefont {van Anders}\ \emph
  {et~al.}(2014{\natexlab{a}})\citenamefont {van Anders}, \citenamefont
  {Ahmed}, \citenamefont {Smith}, \citenamefont {Engel},\ and\ \citenamefont
  {Glotzer}}]{vanAnders2014}%
  \BibitemOpen
  \bibfield  {author} {\bibinfo {author} {\bibfnamefont {G.}~\bibnamefont {van
  Anders}}, \bibinfo {author} {\bibfnamefont {N.~K.}\ \bibnamefont {Ahmed}},
  \bibinfo {author} {\bibfnamefont {R.}~\bibnamefont {Smith}}, \bibinfo
  {author} {\bibfnamefont {M.}~\bibnamefont {Engel}}, \ and\ \bibinfo {author}
  {\bibfnamefont {S.~C.}\ \bibnamefont {Glotzer}},\ }\href {\doibase
  10.1021/nn4057353} {\bibfield  {journal} {\bibinfo  {journal} {ACS Nano}\
  }\textbf {\bibinfo {volume} {8}},\ \bibinfo {pages} {931} (\bibinfo {year}
  {2014}{\natexlab{a}})},\ \bibinfo {note} {pMID: 24359081},\ \Eprint
  {http://arxiv.org/abs/http://dx.doi.org/10.1021/nn4057353}
  {http://dx.doi.org/10.1021/nn4057353} \BibitemShut {NoStop}%
\bibitem [{\citenamefont {van Anders}\ \emph
  {et~al.}(2014{\natexlab{b}})\citenamefont {van Anders}, \citenamefont
  {Klotsa}, \citenamefont {Ahmed}, \citenamefont {Engel},\ and\ \citenamefont
  {Glotzer}}]{Anders2014}%
  \BibitemOpen
  \bibfield  {author} {\bibinfo {author} {\bibfnamefont {G.}~\bibnamefont {van
  Anders}}, \bibinfo {author} {\bibfnamefont {D.}~\bibnamefont {Klotsa}},
  \bibinfo {author} {\bibfnamefont {N.~K.}\ \bibnamefont {Ahmed}}, \bibinfo
  {author} {\bibfnamefont {M.}~\bibnamefont {Engel}}, \ and\ \bibinfo {author}
  {\bibfnamefont {S.~C.}\ \bibnamefont {Glotzer}},\ }\href {\doibase
  10.1073/pnas.1418159111} {\bibfield  {journal} {\bibinfo  {journal} {Proc.
  Natl. Acad. Sci. U.S.A.}\ }\textbf {\bibinfo {volume} {111}},\ \bibinfo
  {pages} {E4812} (\bibinfo {year} {2014}{\natexlab{b}})},\ \Eprint
  {http://arxiv.org/abs/http://www.pnas.org/content/111/45/E4812.full.pdf}
  {http://www.pnas.org/content/111/45/E4812.full.pdf} \BibitemShut {NoStop}%
\bibitem [{\citenamefont {Damasceno}\ \emph
  {et~al.}(2012{\natexlab{b}})\citenamefont {Damasceno}, \citenamefont
  {Engel},\ and\ \citenamefont {Glotzer}}]{Damasceno}%
  \BibitemOpen
  \bibfield  {author} {\bibinfo {author} {\bibfnamefont {P.~F.}\ \bibnamefont
  {Damasceno}}, \bibinfo {author} {\bibfnamefont {M.}~\bibnamefont {Engel}}, \
  and\ \bibinfo {author} {\bibfnamefont {S.~C.}\ \bibnamefont {Glotzer}},\
  }\href {\doibase 10.1021/nn204012y} {\bibfield  {journal} {\bibinfo
  {journal} {ACS Nano}\ }\textbf {\bibinfo {volume} {6}},\ \bibinfo {pages}
  {609} (\bibinfo {year} {2012}{\natexlab{b}})},\ \bibinfo {note} {pMID:
  22098586},\ \Eprint
  {http://arxiv.org/abs/http://dx.doi.org/10.1021/nn204012y}
  {http://dx.doi.org/10.1021/nn204012y} \BibitemShut {NoStop}%
\bibitem [{\citenamefont {Millan}\ \emph {et~al.}(2014)\citenamefont {Millan},
  \citenamefont {Ortiz}, \citenamefont {van Anders},\ and\ \citenamefont
  {Glotzer}}]{Millan2014}%
  \BibitemOpen
  \bibfield  {author} {\bibinfo {author} {\bibfnamefont {J.~A.}\ \bibnamefont
  {Millan}}, \bibinfo {author} {\bibfnamefont {D.}~\bibnamefont {Ortiz}},
  \bibinfo {author} {\bibfnamefont {G.}~\bibnamefont {van Anders}}, \ and\
  \bibinfo {author} {\bibfnamefont {S.~C.}\ \bibnamefont {Glotzer}},\ }\href
  {\doibase 10.1021/nn500147u} {\bibfield  {journal} {\bibinfo  {journal} {ACS
  Nano}\ }\textbf {\bibinfo {volume} {8}},\ \bibinfo {pages} {2918} (\bibinfo
  {year} {2014})},\ \bibinfo {note} {pMID: 24483709},\ \Eprint
  {http://arxiv.org/abs/http://dx.doi.org/10.1021/nn500147u}
  {http://dx.doi.org/10.1021/nn500147u} \BibitemShut {NoStop}%
\bibitem [{\citenamefont {Doppelbauer}\ \emph {et~al.}(2012)\citenamefont
  {Doppelbauer}, \citenamefont {Noya}, \citenamefont {Bianchi},\ and\
  \citenamefont {Kahl}}]{Doppelbauer2012}%
  \BibitemOpen
  \bibfield  {author} {\bibinfo {author} {\bibfnamefont {G.}~\bibnamefont
  {Doppelbauer}}, \bibinfo {author} {\bibfnamefont {E.~G.}\ \bibnamefont
  {Noya}}, \bibinfo {author} {\bibfnamefont {E.}~\bibnamefont {Bianchi}}, \
  and\ \bibinfo {author} {\bibfnamefont {G.}~\bibnamefont {Kahl}},\ }\href
  {\doibase 10.1039/C2SM26043C} {\bibfield  {journal} {\bibinfo  {journal}
  {Soft Matter}\ }\textbf {\bibinfo {volume} {8}},\ \bibinfo {pages} {7768}
  (\bibinfo {year} {2012})}\BibitemShut {NoStop}%
\bibitem [{\citenamefont {Glotzer}\ and\ \citenamefont
  {Solomon}(2007)}]{Glotzer2007}%
  \BibitemOpen
  \bibfield  {author} {\bibinfo {author} {\bibfnamefont {S.~C.}\ \bibnamefont
  {Glotzer}}\ and\ \bibinfo {author} {\bibfnamefont {M.~J.}\ \bibnamefont
  {Solomon}},\ }\href {http://dx.doi.org/10.1038/nmat1949} {\bibfield
  {journal} {\bibinfo  {journal} {Nat Mater}\ }\textbf {\bibinfo {volume}
  {6}},\ \bibinfo {pages} {557} (\bibinfo {year} {2007})}\BibitemShut {NoStop}%
\bibitem [{\citenamefont {Kraft}\ \emph {et~al.}(2011)\citenamefont {Kraft},
  \citenamefont {Hilhorst}, \citenamefont {Heinen}, \citenamefont {Hoogenraad},
  \citenamefont {Luigjes},\ and\ \citenamefont {Kegel}}]{Kraft2011}%
  \BibitemOpen
  \bibfield  {author} {\bibinfo {author} {\bibfnamefont {D.~J.}\ \bibnamefont
  {Kraft}}, \bibinfo {author} {\bibfnamefont {J.}~\bibnamefont {Hilhorst}},
  \bibinfo {author} {\bibfnamefont {M.~A.~P.}\ \bibnamefont {Heinen}}, \bibinfo
  {author} {\bibfnamefont {M.~J.}\ \bibnamefont {Hoogenraad}}, \bibinfo
  {author} {\bibfnamefont {B.}~\bibnamefont {Luigjes}}, \ and\ \bibinfo
  {author} {\bibfnamefont {W.~K.}\ \bibnamefont {Kegel}},\ }\href {\doibase
  10.1021/jp108760g} {\bibfield  {journal} {\bibinfo  {journal} {The Journal of
  Physical Chemistry B}\ }\textbf {\bibinfo {volume} {115}},\ \bibinfo {pages}
  {7175} (\bibinfo {year} {2011})},\ \bibinfo {note} {pMID: 20939541},\ \Eprint
  {http://arxiv.org/abs/http://dx.doi.org/10.1021/jp108760g}
  {http://dx.doi.org/10.1021/jp108760g} \BibitemShut {NoStop}%
\bibitem [{\citenamefont {Kraft}\ \emph {et~al.}(2012)\citenamefont {Kraft},
  \citenamefont {Ni}, \citenamefont {Smallenburg}, \citenamefont {Hermes},
  \citenamefont {Yoon}, \citenamefont {Weitz}, \citenamefont {van Blaaderen},
  \citenamefont {Groenewold}, \citenamefont {Dijkstra},\ and\ \citenamefont
  {Kegel}}]{Kraft2012}%
  \BibitemOpen
  \bibfield  {author} {\bibinfo {author} {\bibfnamefont {D.~J.}\ \bibnamefont
  {Kraft}}, \bibinfo {author} {\bibfnamefont {R.}~\bibnamefont {Ni}}, \bibinfo
  {author} {\bibfnamefont {F.}~\bibnamefont {Smallenburg}}, \bibinfo {author}
  {\bibfnamefont {M.}~\bibnamefont {Hermes}}, \bibinfo {author} {\bibfnamefont
  {K.}~\bibnamefont {Yoon}}, \bibinfo {author} {\bibfnamefont {D.~A.}\
  \bibnamefont {Weitz}}, \bibinfo {author} {\bibfnamefont {A.}~\bibnamefont
  {van Blaaderen}}, \bibinfo {author} {\bibfnamefont {J.}~\bibnamefont
  {Groenewold}}, \bibinfo {author} {\bibfnamefont {M.}~\bibnamefont
  {Dijkstra}}, \ and\ \bibinfo {author} {\bibfnamefont {W.~K.}\ \bibnamefont
  {Kegel}},\ }\href {\doibase 10.1073/pnas.1116820109} {\bibfield  {journal}
  {\bibinfo  {journal} {Proc. Natl. Acad. Sci. U.S.A.}\ }\textbf {\bibinfo
  {volume} {109}},\ \bibinfo {pages} {10787} (\bibinfo {year}
  {2012})}\BibitemShut {NoStop}%
\bibitem [{\citenamefont {Zhang}\ and\ \citenamefont
  {Glotzer}(2004)}]{Zhang2004}%
  \BibitemOpen
  \bibfield  {author} {\bibinfo {author} {\bibfnamefont {Z.}~\bibnamefont
  {Zhang}}\ and\ \bibinfo {author} {\bibfnamefont {S.~C.}\ \bibnamefont
  {Glotzer}},\ }\href {\doibase 10.1021/nl0493500} {\bibfield  {journal}
  {\bibinfo  {journal} {Nano Letters}\ }\textbf {\bibinfo {volume} {4}},\
  \bibinfo {pages} {1407} (\bibinfo {year} {2004})},\ \Eprint
  {http://arxiv.org/abs/http://dx.doi.org/10.1021/nl0493500}
  {http://dx.doi.org/10.1021/nl0493500} \BibitemShut {NoStop}%
\bibitem [{\citenamefont {Donev}\ \emph {et~al.}(2006)\citenamefont {Donev},
  \citenamefont {Burton}, \citenamefont {Stillinger},\ and\ \citenamefont
  {Torquato}}]{Donev2006}%
  \BibitemOpen
  \bibfield  {author} {\bibinfo {author} {\bibfnamefont {A.}~\bibnamefont
  {Donev}}, \bibinfo {author} {\bibfnamefont {J.}~\bibnamefont {Burton}},
  \bibinfo {author} {\bibfnamefont {F.~H.}\ \bibnamefont {Stillinger}}, \ and\
  \bibinfo {author} {\bibfnamefont {S.}~\bibnamefont {Torquato}},\ }\href
  {\doibase 10.1103/PhysRevB.73.054109} {\bibfield  {journal} {\bibinfo
  {journal} {Phys. Rev. B}\ }\textbf {\bibinfo {volume} {73}},\ \bibinfo
  {pages} {054109} (\bibinfo {year} {2006})}\BibitemShut {NoStop}%
\bibitem [{\citenamefont {Avendano}\ and\ \citenamefont
  {Escobedo}(2012)}]{Avendano2012}%
  \BibitemOpen
  \bibfield  {author} {\bibinfo {author} {\bibfnamefont {C.}~\bibnamefont
  {Avendano}}\ and\ \bibinfo {author} {\bibfnamefont {F.~A.}\ \bibnamefont
  {Escobedo}},\ }\href {\doibase 10.1039/C2SM07428A} {\bibfield  {journal}
  {\bibinfo  {journal} {Soft Matter}\ }\textbf {\bibinfo {volume} {8}},\
  \bibinfo {pages} {4675} (\bibinfo {year} {2012})}\BibitemShut {NoStop}%
\bibitem [{\citenamefont {Smallenburg}\ \emph {et~al.}(2012)\citenamefont
  {Smallenburg}, \citenamefont {Filion}, \citenamefont {Marechal},\ and\
  \citenamefont {Dijkstra}}]{Smallenburg2012}%
  \BibitemOpen
  \bibfield  {author} {\bibinfo {author} {\bibfnamefont {F.}~\bibnamefont
  {Smallenburg}}, \bibinfo {author} {\bibfnamefont {L.}~\bibnamefont {Filion}},
  \bibinfo {author} {\bibfnamefont {M.}~\bibnamefont {Marechal}}, \ and\
  \bibinfo {author} {\bibfnamefont {M.}~\bibnamefont {Dijkstra}},\ }\href
  {\doibase 10.1073/pnas.1211784109} {\bibfield  {journal} {\bibinfo  {journal}
  {Proc. Natl. Acad. Sci. U.S.A.}\ }\textbf {\bibinfo {volume} {109}},\
  \bibinfo {pages} {17886} (\bibinfo {year} {2012})},\ \Eprint
  {http://arxiv.org/abs/http://www.pnas.org/content/109/44/17886.full.pdf}
  {http://www.pnas.org/content/109/44/17886.full.pdf} \BibitemShut {NoStop}%
\bibitem [{\citenamefont {Wojciechowski}\ and\ \citenamefont
  {Frenkel}(2004)}]{Wojciechowski2004}%
  \BibitemOpen
  \bibfield  {author} {\bibinfo {author} {\bibfnamefont {K.~W.}\ \bibnamefont
  {Wojciechowski}}\ and\ \bibinfo {author} {\bibfnamefont {D.}~\bibnamefont
  {Frenkel}},\ }\href@noop {} {\bibfield  {journal} {\bibinfo  {journal}
  {Computational Methods in Science and Technology}\ }\textbf {\bibinfo
  {volume} {10}},\ \bibinfo {pages} {235} (\bibinfo {year} {2004})}\BibitemShut
  {NoStop}%
\bibitem [{\citenamefont {Zhao}\ \emph {et~al.}(2011)\citenamefont {Zhao},
  \citenamefont {Bruinsma},\ and\ \citenamefont {Mason}}]{Zhao2011}%
  \BibitemOpen
  \bibfield  {author} {\bibinfo {author} {\bibfnamefont {K.}~\bibnamefont
  {Zhao}}, \bibinfo {author} {\bibfnamefont {R.}~\bibnamefont {Bruinsma}}, \
  and\ \bibinfo {author} {\bibfnamefont {T.~G.}\ \bibnamefont {Mason}},\ }\href
  {\doibase 10.1073/pnas.1014942108} {\bibfield  {journal} {\bibinfo  {journal}
  {Proc. Natl. Acad. Sci. U.S.A.}\ }\textbf {\bibinfo {volume} {108}},\
  \bibinfo {pages} {2684} (\bibinfo {year} {2011})}\BibitemShut {NoStop}%
\bibitem [{\citenamefont {John}\ and\ \citenamefont
  {Escobedo}(2005)}]{John2005}%
  \BibitemOpen
  \bibfield  {author} {\bibinfo {author} {\bibfnamefont {B.~S.}\ \bibnamefont
  {John}}\ and\ \bibinfo {author} {\bibfnamefont {F.~A.}\ \bibnamefont
  {Escobedo}},\ }\href {\doibase 10.1021/jp0551521} {\bibfield  {journal}
  {\bibinfo  {journal} {The Journal of Physical Chemistry B}\ }\textbf
  {\bibinfo {volume} {109}},\ \bibinfo {pages} {23008} (\bibinfo {year}
  {2005})},\ \Eprint {http://arxiv.org/abs/http://dx.doi.org/10.1021/jp0551521}
  {http://dx.doi.org/10.1021/jp0551521} \BibitemShut {NoStop}%
\bibitem [{\citenamefont {Triplett}\ and\ \citenamefont
  {Fichthorn}(2008)}]{Triplett2008}%
  \BibitemOpen
  \bibfield  {author} {\bibinfo {author} {\bibfnamefont {D.~A.}\ \bibnamefont
  {Triplett}}\ and\ \bibinfo {author} {\bibfnamefont {K.~A.}\ \bibnamefont
  {Fichthorn}},\ }\href {\doibase 10.1103/PhysRevE.77.011707} {\bibfield
  {journal} {\bibinfo  {journal} {Phys. Rev. E}\ }\textbf {\bibinfo {volume}
  {77}},\ \bibinfo {pages} {011707} (\bibinfo {year} {2008})}\BibitemShut
  {NoStop}%
\bibitem [{\citenamefont {Anderson}\ \emph {et~al.}(2013)\citenamefont
  {Anderson}, \citenamefont {Jankowski}, \citenamefont {Grubb}, \citenamefont
  {Engel},\ and\ \citenamefont {Glotzer}}]{Anderson2013}%
  \BibitemOpen
  \bibfield  {author} {\bibinfo {author} {\bibfnamefont {J.~A.}\ \bibnamefont
  {Anderson}}, \bibinfo {author} {\bibfnamefont {E.}~\bibnamefont {Jankowski}},
  \bibinfo {author} {\bibfnamefont {T.~L.}\ \bibnamefont {Grubb}}, \bibinfo
  {author} {\bibfnamefont {M.}~\bibnamefont {Engel}}, \ and\ \bibinfo {author}
  {\bibfnamefont {S.~C.}\ \bibnamefont {Glotzer}},\ }\href {\doibase
  10.1016/j.jcp.2013.07.023} {\bibfield  {journal} {\bibinfo  {journal}
  {Journal of Computational Physics}\ }\textbf {\bibinfo {volume} {254}},\
  \bibinfo {pages} {27} (\bibinfo {year} {2013})}\BibitemShut {NoStop}%
\bibitem [{\citenamefont {Anderson}\ \emph {et~al.}(2015)\citenamefont
  {Anderson}, \citenamefont {Irrgang},\ and\ \citenamefont
  {Glotzer}}]{Anderson2015}%
  \BibitemOpen
  \bibfield  {author} {\bibinfo {author} {\bibfnamefont {J.~A.}\ \bibnamefont
  {Anderson}}, \bibinfo {author} {\bibfnamefont {M.~E.}\ \bibnamefont
  {Irrgang}}, \ and\ \bibinfo {author} {\bibfnamefont {S.~C.}\ \bibnamefont
  {Glotzer}},\ }\href@noop {} {\bibfield  {journal} {\bibinfo  {journal}
  {Preprint}\ } (\bibinfo {year} {2015})}\BibitemShut {NoStop}%
\bibitem [{hoo()}]{hoomd}%
  \BibitemOpen
  \href {http://codeblue.umich.edu/hoomd-blue} {\enquote {\bibinfo {title}
  {Hoomd-blue},}\ }\bibinfo {howpublished}
  {\url{http://codeblue.umich.edu/hoomd-blue}}\BibitemShut {NoStop}%
\bibitem [{\citenamefont {Anderson}\ \emph {et~al.}(2008)\citenamefont
  {Anderson}, \citenamefont {Lorenz},\ and\ \citenamefont
  {Travesset}}]{Anderson2008}%
  \BibitemOpen
  \bibfield  {author} {\bibinfo {author} {\bibfnamefont {J.~A.}\ \bibnamefont
  {Anderson}}, \bibinfo {author} {\bibfnamefont {C.~D.}\ \bibnamefont
  {Lorenz}}, \ and\ \bibinfo {author} {\bibfnamefont {A.}~\bibnamefont
  {Travesset}},\ }\href {\doibase 10.1016/j.jcp.2008.01.047} {\bibfield
  {journal} {\bibinfo  {journal} {J. Comp. Phys.}\ }\textbf {\bibinfo {volume}
  {227}},\ \bibinfo {pages} {5342 } (\bibinfo {year} {2008})},\ \bibinfo {note}
  {{\url{http://codeblue.umich.edu/hoomd-blue}}}\BibitemShut {NoStop}%
\bibitem [{\citenamefont {Towns}\ \emph {et~al.}(2014)\citenamefont {Towns},
  \citenamefont {Cockerill}, \citenamefont {Dahan}, \citenamefont {Foster},
  \citenamefont {Gaither}, \citenamefont {Grimshaw}, \citenamefont {Hazlewood},
  \citenamefont {Lathrop}, \citenamefont {Lifka}, \citenamefont {Peterson},
  \citenamefont {Roskies}, \citenamefont {Scott},\ and\ \citenamefont
  {Wilkens-Diehr}}]{XSEDE}%
  \BibitemOpen
  \bibfield  {author} {\bibinfo {author} {\bibfnamefont {J.}~\bibnamefont
  {Towns}}, \bibinfo {author} {\bibfnamefont {T.}~\bibnamefont {Cockerill}},
  \bibinfo {author} {\bibfnamefont {M.}~\bibnamefont {Dahan}}, \bibinfo
  {author} {\bibfnamefont {I.}~\bibnamefont {Foster}}, \bibinfo {author}
  {\bibfnamefont {K.}~\bibnamefont {Gaither}}, \bibinfo {author} {\bibfnamefont
  {A.}~\bibnamefont {Grimshaw}}, \bibinfo {author} {\bibfnamefont
  {V.}~\bibnamefont {Hazlewood}}, \bibinfo {author} {\bibfnamefont
  {S.}~\bibnamefont {Lathrop}}, \bibinfo {author} {\bibfnamefont
  {D.}~\bibnamefont {Lifka}}, \bibinfo {author} {\bibfnamefont {G.~D.}\
  \bibnamefont {Peterson}}, \bibinfo {author} {\bibfnamefont {R.}~\bibnamefont
  {Roskies}}, \bibinfo {author} {\bibfnamefont {J.~R.}\ \bibnamefont {Scott}},
  \ and\ \bibinfo {author} {\bibfnamefont {N.}~\bibnamefont {Wilkens-Diehr}},\
  }\href {\doibase http://doi.ieeecomputersociety.org/10.1109/MCSE.2014.80}
  {\bibfield  {journal} {\bibinfo  {journal} {Computing in Science and
  Engineering}\ }\textbf {\bibinfo {volume} {16}},\ \bibinfo {pages} {62}
  (\bibinfo {year} {2014})}\BibitemShut {NoStop}%
\bibitem [{\citenamefont {of~Texas~at Austin}()}]{stampede}%
  \BibitemOpen
  \bibfield  {author} {\bibinfo {author} {\bibfnamefont {T.~U.}\ \bibnamefont
  {of~Texas~at Austin}},\ }\href {http://www.tacc.utexas.edu} {\enquote
  {\bibinfo {title} {Texas advanced computing center},}\ }\BibitemShut
  {NoStop}%
\bibitem [{\citenamefont {Glaser}\ \emph {et~al.}(2015)\citenamefont {Glaser},
  \citenamefont {Nguyen}, \citenamefont {Anderson}, \citenamefont {Lui},
  \citenamefont {Spiga}, \citenamefont {Millan}, \citenamefont {Morse},\ and\
  \citenamefont {Glotzer}}]{Glaser2015}%
  \BibitemOpen
  \bibfield  {author} {\bibinfo {author} {\bibfnamefont {J.}~\bibnamefont
  {Glaser}}, \bibinfo {author} {\bibfnamefont {T.~D.}\ \bibnamefont {Nguyen}},
  \bibinfo {author} {\bibfnamefont {J.~A.}\ \bibnamefont {Anderson}}, \bibinfo
  {author} {\bibfnamefont {P.}~\bibnamefont {Lui}}, \bibinfo {author}
  {\bibfnamefont {F.}~\bibnamefont {Spiga}}, \bibinfo {author} {\bibfnamefont
  {J.~A.}\ \bibnamefont {Millan}}, \bibinfo {author} {\bibfnamefont {D.~C.}\
  \bibnamefont {Morse}}, \ and\ \bibinfo {author} {\bibfnamefont {S.~C.}\
  \bibnamefont {Glotzer}},\ }\href {\doibase 10.1016/j.cpc.2015.02.028}
  {\bibfield  {journal} {\bibinfo  {journal} {Computer Physics Communications}\
  }\textbf {\bibinfo {volume} {192}},\ \bibinfo {pages} {97} (\bibinfo {year}
  {2015})}\BibitemShut {NoStop}%
\bibitem [{\citenamefont {Bernard}\ and\ \citenamefont
  {Krauth}(2011)}]{Bernard2011}%
  \BibitemOpen
  \bibfield  {author} {\bibinfo {author} {\bibfnamefont {E.~P.}\ \bibnamefont
  {Bernard}}\ and\ \bibinfo {author} {\bibfnamefont {W.}~\bibnamefont
  {Krauth}},\ }\href {\doibase 10.1103/PhysRevLett.107.155704} {\bibfield
  {journal} {\bibinfo  {journal} {Phys. Rev. Lett.}\ }\textbf {\bibinfo
  {volume} {107}},\ \bibinfo {pages} {155704} (\bibinfo {year}
  {2011})}\BibitemShut {NoStop}%
\bibitem [{\citenamefont {Deutschl\"ander}\ \emph {et~al.}(2013)\citenamefont
  {Deutschl\"ander}, \citenamefont {Horn}, \citenamefont {L\"owen},
  \citenamefont {Maret},\ and\ \citenamefont {Keim}}]{Deutschlander2013}%
  \BibitemOpen
  \bibfield  {author} {\bibinfo {author} {\bibfnamefont {S.}~\bibnamefont
  {Deutschl\"ander}}, \bibinfo {author} {\bibfnamefont {T.}~\bibnamefont
  {Horn}}, \bibinfo {author} {\bibfnamefont {H.}~\bibnamefont {L\"owen}},
  \bibinfo {author} {\bibfnamefont {G.}~\bibnamefont {Maret}}, \ and\ \bibinfo
  {author} {\bibfnamefont {P.}~\bibnamefont {Keim}},\ }\href {\doibase
  10.1103/PhysRevLett.111.098301} {\bibfield  {journal} {\bibinfo  {journal}
  {Phys. Rev. Lett.}\ }\textbf {\bibinfo {volume} {111}},\ \bibinfo {pages}
  {098301} (\bibinfo {year} {2013})}\BibitemShut {NoStop}%
\bibitem [{\citenamefont {Dillmann}\ \emph {et~al.}(2012)\citenamefont
  {Dillmann}, \citenamefont {Maret},\ and\ \citenamefont
  {Keim}}]{Dillmann2012}%
  \BibitemOpen
  \bibfield  {author} {\bibinfo {author} {\bibfnamefont {P.}~\bibnamefont
  {Dillmann}}, \bibinfo {author} {\bibfnamefont {G.}~\bibnamefont {Maret}}, \
  and\ \bibinfo {author} {\bibfnamefont {P.}~\bibnamefont {Keim}},\ }\href
  {http://stacks.iop.org/0953-8984/24/i=46/a=464118} {\bibfield  {journal}
  {\bibinfo  {journal} {Journal of Physics: Condensed Matter}\ }\textbf
  {\bibinfo {volume} {24}},\ \bibinfo {pages} {464118} (\bibinfo {year}
  {2012})}\BibitemShut {NoStop}%
\bibitem [{\citenamefont {Schultz}\ \emph {et~al.}(2015)\citenamefont
  {Schultz}, \citenamefont {Damasceno}, \citenamefont {Engel},\ and\
  \citenamefont {Glotzer}}]{Schultz2015}%
  \BibitemOpen
  \bibfield  {author} {\bibinfo {author} {\bibfnamefont {B.~A.}\ \bibnamefont
  {Schultz}}, \bibinfo {author} {\bibfnamefont {P.~F.}\ \bibnamefont
  {Damasceno}}, \bibinfo {author} {\bibfnamefont {M.}~\bibnamefont {Engel}}, \
  and\ \bibinfo {author} {\bibfnamefont {S.~C.}\ \bibnamefont {Glotzer}},\
  }\href {\doibase 10.1021/nn507490j} {\bibfield  {journal} {\bibinfo
  {journal} {ACS Nano}\ }\textbf {\bibinfo {volume} {9}},\ \bibinfo {pages}
  {2336} (\bibinfo {year} {2015})},\ \bibinfo {note} {pMID: 25692863},\ \Eprint
  {http://arxiv.org/abs/http://dx.doi.org/10.1021/nn507490j}
  {http://dx.doi.org/10.1021/nn507490j} \BibitemShut {NoStop}%
\bibitem [{\citenamefont {Gantapara}\ \emph {et~al.}(2015)\citenamefont
  {Gantapara}, \citenamefont {Qi},\ and\ \citenamefont
  {Dijkstra}}]{Gantapara2015}%
  \BibitemOpen
  \bibfield  {author} {\bibinfo {author} {\bibfnamefont {A.~P.}\ \bibnamefont
  {Gantapara}}, \bibinfo {author} {\bibfnamefont {W.}~\bibnamefont {Qi}}, \
  and\ \bibinfo {author} {\bibfnamefont {M.}~\bibnamefont {Dijkstra}},\ }\href
  {http://arxiv.org/abs/1504.03130} {\bibfield  {journal} {\bibinfo  {journal}
  {arXiv}\ } (\bibinfo {year} {2015})}\BibitemShut {NoStop}%
\bibitem [{\citenamefont {Gao}\ \emph {et~al.}(2013)\citenamefont {Gao},
  \citenamefont {Alvi}, \citenamefont {Rosen}, \citenamefont {Lav},\ and\
  \citenamefont {Tao}}]{Gao2013}%
  \BibitemOpen
  \bibfield  {author} {\bibinfo {author} {\bibfnamefont {B.}~\bibnamefont
  {Gao}}, \bibinfo {author} {\bibfnamefont {Y.}~\bibnamefont {Alvi}}, \bibinfo
  {author} {\bibfnamefont {D.}~\bibnamefont {Rosen}}, \bibinfo {author}
  {\bibfnamefont {M.}~\bibnamefont {Lav}}, \ and\ \bibinfo {author}
  {\bibfnamefont {A.~R.}\ \bibnamefont {Tao}},\ }\href {\doibase
  10.1039/C2CC37158H} {\bibfield  {journal} {\bibinfo  {journal} {Chem.
  Commun.}\ }\textbf {\bibinfo {volume} {49}},\ \bibinfo {pages} {4382}
  (\bibinfo {year} {2013})}\BibitemShut {NoStop}%
\bibitem [{\citenamefont {Rosen}\ and\ \citenamefont {Tao}(2014)}]{Rosen2014}%
  \BibitemOpen
  \bibfield  {author} {\bibinfo {author} {\bibfnamefont {D.~A.}\ \bibnamefont
  {Rosen}}\ and\ \bibinfo {author} {\bibfnamefont {A.~R.}\ \bibnamefont
  {Tao}},\ }\href {\doibase 10.1021/am4057612} {\bibfield  {journal} {\bibinfo
  {journal} {ACS Applied Materials \& Interfaces}\ }\textbf {\bibinfo {volume}
  {6}},\ \bibinfo {pages} {4134} (\bibinfo {year} {2014})},\ \bibinfo {note}
  {pMID: 24533909},\ \Eprint
  {http://arxiv.org/abs/http://dx.doi.org/10.1021/am4057612}
  {http://dx.doi.org/10.1021/am4057612} \BibitemShut {NoStop}%
\end{thebibliography}%
% \bibliography{bib.bib}
\bibliographystyle{apsrev4-1}

\end{document}